# Entanglement in a Jaynes-Cummings Model with Two Atoms and Two Photon Modes

Samina S. Masood[1,1†] and Allen Miller[2,*]

[1]Department of Physics, University of Houston Clear Lake, Houston TX 77058
[2]Department of Physics, Syracuse University, Syracuse, NY 13244-1130

## Abstract

We investigate the conditions of entanglement for a system of two atoms and two photon modes in vacuum, using the Jaynes-Cummings model in the rotating-wave approximation. It is found, by generalizing the existing results, that the strength of entanglement is a periodic function of time. We explicitly show that our results are in agreement with the existing results of entanglement conditions under appropriate limits. Results for the two-atom and two-photon system are generalized to the case of arbitrary values for the atomic energies, corresponding to photon modes frequencies. Though it is apparently a generalization of the existing work, we have considered for the first time both the resonant and non-resonant conditions and found a general equation which could be true for both cases. Moreover, we show that periodicity of the entanglement is a distinct feature of resonant system. Considering the two atoms and two photons system, in detail, we setup an approach which could be generalized for many particle systems and the resulting master equation can also be analyzed.

## Introduction

Entanglement of quantum states is not a new concept; however it is not a property of Fock Space [1]. Therefore, it does not appear automatically in a vacuum and one has to develop a special representation using second quantization to entangle atomic states with vacuum. This phenomenon is still not well-understood. The possibility of entanglement in the second quantization [1-4], using simple theoretical models has not been understood in detail yet. Most of the existing literature on entanglement in the second quantization will be reviewed in this paper and we will compare our results with them.

Pawlowski and Czachor (PC) in Ref.[1] have used a simple model in a system with two atoms and two photon modes. They found that the entanglement of two atoms with the vacuum can occur using the canonical commutation relations. On the other hand, the Jaynes and Cummings (JC) model [5] is considered to be one of the most appropriate models for the purpose of analyzing ion traps in cavities. The JC model, being a nonlinear model, gives a good theoretical tool to study ion trapping in a cavity using quantum electrodynamics. Hussin and Nieto (HN) in Ref.[6] have studied the JC model in the rotating wave approximation (RWA) to construct coherent quantum states using ladder operators. We use the same model in the same RWA to study entanglement of more than one atom with more than one modes of photon in vacuum. For this purpose we use a form of the JC Hamiltonian used by HN and several other researchers. The stationary states of the JC model in the RWA are given in Ref. [6].

[†]masood@uhcl.edu,
[*]allenmil@syr.edu

The master equation for the cavity losses, using JC Hamiltonian $H_j$ can be written as [7]

$$\frac{d\rho_j}{dt} = -\frac{i}{\hbar}[H_j, \rho_j] + \gamma\left(a_j \rho_j a_j^\dagger - \frac{1}{2} a_j^\dagger a_j \rho_j - \frac{1}{2} \rho_j a_j^\dagger a_j\right),$$

for system of j number of particles. In this paper we study a system with two atoms labeled A and B. We also use two distinct photon modes, labeled as A and B as well. Atom j interacts with mode j only. $\rho_j$ is the density matrix of the atom-cavity system for the jth mode and jth atom. The factor $\gamma$ in the second term is the rate of loss of photons from the cavity, due to imperfect reflectivity of the cavity mirrors. $H_j$ is the JC Hamiltonian for the jth particle, given as:

$$H_j = \hbar\omega_j(N_j + \frac{1}{2})I + \frac{1}{2}E_j \sigma_{zj} + \hbar\kappa_j(a_j^+ \sigma_{-j} + a_j \sigma_{+j}). \quad (1)$$

I is the identity matrix. The Hamiltonian, Eq. (1), is identical to that used in Ref. [7], except that we have included the zero-point energy contribution to the photon energy.

In Eq. (1), each atom, (j = A, B for atoms A and B) has a ground state $|->_j$ and an excited

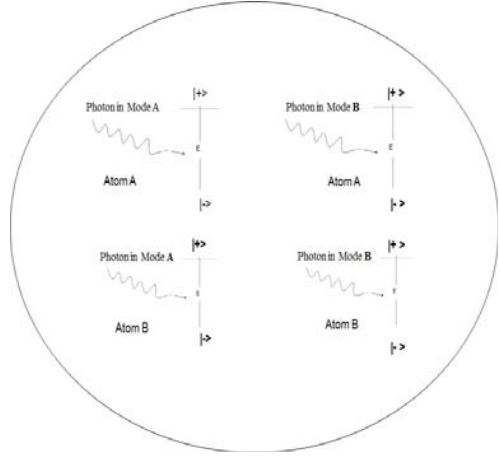

*[Figure 1: Absorption of a photon in an atom in a system of two atoms A&B and two photons in corresponding modes A&B. Atoms can go to an excited state due to the absorption of electron and come back to the ground state by emission of the corresponding mode photon.]*

state $|+>_j$ as shown in Figure 1. The atomic energy difference for atom j is $E_j$. The frequency of photon mode j is denoted as $\omega_j$. The strength of the interaction between atom j and mode j is $\kappa_j$. $\sigma_\pm$ are the Pauli matrices in standard notation. The Pauli matrices act on the atomic states exactly as the Pauli matrices act on a spin one-half particle, with the lower row of the spinor representing the atom's ground state, and the upper row representing the excited state.

We also adopt the usual definitions of the raising and lowering operators: $\sigma_+ = \sigma_x + i\sigma_y$ and $\sigma_- = \sigma_x - i\sigma_y$. The operators for the photon mode are defined in the usual way. The operator $a_j$ is the destruction operator for photons in mode j. Also, $N_j = a_j^\dagger a_j$ is the number operator for mode j. The choice for the zero of energy of atom j is taken to be midway between its ground state and excited state energies.
Moreover,

$$\sigma_{-j} = |->_j <+|_j$$
$$\sigma_{+j} = |+>_j <-|_j$$
$$\sigma_{zj} = |+>_j <+|_j - |->_j <-|_j \qquad (2)$$

The operator $\sigma_{zj}$ acts on atomic states of atom j as the Pauli spin matrix in the z direction does, with the excited state considered as the `up-spin state' and the ground spin state is taken as the `down-spin state'. Then the second term of Eq.(1) denotes the unperturbed atomic state energies. The first term of Eq.(1) is the unperturbed photon energy $\hbar\omega_j$ of mode j. $E_j$ is the change of atomic energy due to the absorption of a photon in mode j.

Our main goal is to study the probability of entanglement if a photon in mode A or B is absorbed or emitted by an atom in state $|->$ or $|+>$, respectively. These atoms may or may not be identical. We shall specialize to the case of identical atoms later, for the sake of simplicity. We use a straightforward generalization of the Hamiltonian used by HN[6], extending it to a two-atom, two photon mode system. The complete Hamiltonian is a sum over j (j = A, B) of Eq. (1). We present our calculations of the stationary states in Section II. In Section III, we study the time evolution of quantum states. Section IV compares our results with those of PC. Finally, section V is devoted to the discussion of results and of possible technical applications.

**STATIONARY STATES**

We start our calculations with the Hamiltonian of Eq. (1) for the system j, (j = A, B), which represents the JC model in the RWA. Each photon mode j can cause a transition of atom j between its ground state $|->$ and its excited state $|+>$ via the emission or absorption of a photon. It can be shown in a straight-forward way that the total number of excitations in the cavity-atom system is given by

$$\mathcal{N}_j = a_j^+ a_j + \frac{\sigma_{zj}}{2} + \frac{1}{2} \qquad (3)$$

This is a constant of motion for the jth mode. From this, one can easily obtain the eigenvalues and eigenstates [7]. The Hamiltonian H for the two atoms, two modes problem can be written as

$$H = H_A + H_B \qquad (4)$$

In writing Eq. (4), it is assumed that the two photon modes A and B are distinct. It can be noted that the model expressed by Eq.(4) could be extended to include an interaction between atom i and photon mode j for $j \neq i$. This would be a generalization of the problem studied in references 1 and 2. It is also worth-mentioning here that Eqs. (1) and (4) contain the unperturbed atom and mode energies ( first two terms of Eq.(1)), as well as the interactions. In the study of PC, the Hamiltonian includes only the interaction terms, the last term in our Eq.(1). Following

HN [6], we introduce the dimensionless parameters $\lambda_j$ and $\varepsilon_j$ by

$$\hbar\omega_j = E_j(1+\varepsilon_j) \tag{5}$$

and

$$\lambda_j = \frac{\hbar\kappa_j}{E_j} \tag{6}$$

The parameter $\varepsilon_j$ is a ``detuning parameter" in that it is a measure of the deviation of the photon energy $\hbar\omega_j$ from the atomic energy difference $E_j$. Then Eq.(1) can be rewritten as:

$$H_j = (1+\varepsilon_j)E_j(N_j + \frac{1}{2}) + \frac{E_j\sigma_{zj}}{2} + \lambda_j E_j(a_j^+\sigma_{-j} + a_j\sigma_{+j}) \tag{7}$$

The lack of coupling between Hamiltonians $H_A$ and $H_B$ means that a complete set of stationary states of H can be formed from products of the stationary states of $H_A$ and $H_B$. The ground state of $H_j$ is simply

$$|G>_j = |0;->_j \tag{8}$$

where $|0;->_j$ denotes the state in which the photon state is the vacuum and atom j is in its ground state. Its energy is $(E_G)_j = \varepsilon_j E_j/2$.

The normalized excited states of $H_j$ can be enumerated by n= 0, 1, 2, ..... They are

$$|\psi_n^->_j = (\cos\theta_{nj})|n;+>_j - (\sin\theta_{nj})|n+1;->_j \tag{9}$$

and also

$$|\psi_n^+>_j = (\sin\theta_{nj})|n;+>_j + (\cos\theta_{nj})|n+1;->_j \tag{10}$$

with energies

$$E_{nj}^{\pm} = (1+\varepsilon_j)(n+1)E_j \pm q_{nj}(n+1)E_j \tag{11}$$

In equations (9) and (10), $|n;\pm>_j$ denotes the state with atom j in atomic state $|\pm>$ and with n photons in mode j. The angle $\theta_{nj}$, appearing in Eqs. (9) and (10) is defined by

$$\cos\theta_{nj} = \sqrt{(q_{n+1,j} + \frac{\varepsilon_j}{2})/2q_{n+1,j}} \tag{12}$$

$$\sin\theta_{nj} = (\frac{\lambda_j}{|\lambda_j|})\sqrt{(q_{n+1,j} - \frac{\varepsilon_j}{2})/2q_{n+1,j}} \tag{13}$$

Finally, $q_{n,j}$ is defined by

$$q_{n,j} = \sqrt{(\frac{\varepsilon_j^2}{4}) + n\lambda_j^2} \tag{14}$$

To write down a basis of stationary states for the full Hamiltonian H, we only need to take products of the stationary states of systems A and B. Then, the ground state of H is

$$|G> = |G>_A |G>_B$$
$$= |0;->_A |0;->_B \qquad (15)$$

The excited states are

$$|\psi_{n(A)}^{\pm}>_A |G>_B, \dots\dots\dots\dots\dots\dots (a)$$

$$|G>_A |\psi_{n(B)}^{\pm}>_B, \dots\dots\dots\dots\dots\dots (b)$$

and

$$|\psi_{n(A)}^{\pm}>_A |\psi_{n(B)}^{\pm}>_B, \dots\dots\dots\dots\dots (c) \qquad (16)$$

with n(A) and n(B) each taking values 0,1,2,...... In Eq. (16c), all four choices of the signs + and - must be included. The problem studied by PC focuses on the vector space V spanned by the four states

$$|\Phi_1> = |0;->_A |1;->_B$$
$$|\Phi_2> = |1;->_A |0;->_B$$
$$|\Phi_3> = |0;+>_A |0;->_B$$
$$|\Phi_4> = |0;->_A |0;+>_B \qquad (17)$$

These four states of Eq.(17) are the tensor product of atomic ground states and excited states with known photon modes. These states can ultimately show entanglement. The study of PC considers the choice of the initial state (t=0) as

$$|\psi_\alpha> = (\frac{1}{\sqrt{2}})(|\Phi_1> + |\Phi_2>) \qquad (18)$$

and the time development of $|\psi_\alpha>$ is analyzed. PC has studied the resonant case ($\varepsilon = 0$). They have only employed the interaction term of the JC Hamiltonian (Eq.(1)). We have included the non-resonant case in the next section also.

**Time Evolution in JC Model**

**General Results**

Eqs.(15) and (16) give the stationary states for the system of two atoms and two photon modes. Inspection of this equation shows that the vector space V is also spanned by the following four stationary states:

$$|\psi_1> = |G>_A |\psi_0^->_B$$
$$|\psi_2> = |\psi_0^->_A |G>_B$$
$$|\psi_3> = |G>_A |\psi_0^+>_B$$
$$|\psi_4> = |\psi_0^+>_A |G>_B \qquad (19)$$

If the initial state is any state in V, its evolution is found by expansion of the initial state in the set $|\psi_k>$, where k=1, 2, 3, and 4. If each term in the expension is multiplied by $\exp{-iE_k t/\hbar}$, (with $E_k$ equal to the energy of $|\psi_k>$), we have the evolution of the initial state.

The energies $E_k$ can be obtained from Eq.(11) by adding the energies of systems A and B. The results are

$$E_1 = E_{GA} + E_{0B}^- = \frac{\varepsilon_A E_A}{2} + (1+\varepsilon_B)E_B - q_{0B}E_B$$

$$E_2 = E_{GB} + E_{0A}^- = \frac{\varepsilon_B E_B}{2} + (1+\varepsilon_A)E_A - q_{0A}E_A$$

$$E_3 = E_{GA} + E_{0B}^+ = \frac{\varepsilon_A E_A}{2} + (1+\varepsilon_B)E_B + q_{0B}E_B$$

$$E_4 = E_{GB} + E_{0A}^+ = \frac{\varepsilon_B E_B}{2} + (1+\varepsilon_A)E_A + q_{0A}E_A$$

(20)

We now consider the case for which the initial state is given by Eq. (18). The expansion of $|\psi_\alpha>$ in the states $|\psi_m>$ is

$$|\psi_\alpha> = \frac{1}{\sqrt{2}}\sum_{m=1}^{4} c_m |\psi_m>$$

(21)

Where,

$$c_1 = -\sin\theta_B; \quad\quad\quad\quad\quad(a)$$
$$c_2 = -\sin\theta_A; \quad\quad\quad\quad\quad(b)$$
$$c_3 = \cos\theta_B; \quad\quad\quad\quad\quad\quad(c)$$
$$c_4 = \cos\theta_A \quad\quad\quad\quad\quad\quad(d)$$

(22)

In Eq.(22), $\theta_{0A}$ and $\theta_{0B}$ are replaced by $\theta_A$ and $\theta_B$ respectively. If $|\psi_\alpha>$ in Eq.(19) is the full system state at time t=0, then its evolution is given by

$$|\psi_\alpha(t)> = \frac{1}{\sqrt{2}}\sum_{k=1}^{4} c_k \exp(-iE_k t/\hbar)|\psi_k>$$

(23)

Eq.(23) gives the general case of time evolution. We can study it particularly for our proposed system as a special case and discuss the pattern of superposition of wave functions.

**Special Case**

To analyze the time development, first consider the special case of $\kappa_A = \kappa_B = \kappa$, $E_A = E_B = E_{atom},$ and $\omega_A = \omega_B = \omega.$ Then, $\varepsilon_A = \varepsilon_B = \varepsilon$. It also should be noted that then the two photon modes have the same frequency. Since we have assumed that the two photon modes are distinct, the two polarization directions of the mode must be perpendicular. Also note that we are not necessarily at resonance, i.e., $\varepsilon$ is not necessarily equal to zero.

Continuing, for this special case, we also have $q_A = q_B = q$ with

$$q = \sqrt{\lambda^2 + (\frac{\varepsilon}{4})^2}$$

(24)

and $\theta_A = \theta_B = \theta$ where $\theta$ is given by

$$\cos\theta = \frac{1}{\sqrt{2}}\sqrt{1+(\frac{\varepsilon}{2q})}$$
$$\sin\theta = \frac{1}{\sqrt{2}}\frac{\lambda}{|\lambda|}\sqrt{1-(\frac{\varepsilon}{2q})}$$
(25)

The time evolution of state of the full system is then

$$|\psi_\alpha(t)> = \frac{1}{\sqrt{2}}\{-\sin\theta\exp(-iE_1 t/\hbar)[|\psi_1> + |\psi_2>] + \cos\theta\exp(-iE_3 t/\hbar)[|\psi_3> + |\psi_4>]\}$$
(26)

To obtain Eq.(26), we have made use of the fact that $E_2 = E_1$ and $E_4 = E_3$, for this special case. The values of $E_1$ and $E_3$ are

$$E_1 = E_{atom}\left[1+\frac{3}{2}\varepsilon - q\right],$$
$$E_3 = E_{atom}\left[1+\frac{3}{2}\varepsilon + q\right]$$
(27)

We have not yet assumed the resonant behavior, when the detuning parameter is taken to be zero.

**Resonant Subcase ( $\varepsilon = 0$ )**

The evolution of $|\psi_\alpha(t)>$ is particularly simple for the subcase in which the detuning parameter $\varepsilon = 0$. Then, Eq. (25) yields

$$\cos\theta = \frac{1}{\sqrt{2}};$$
(28)

$$\sin\theta = \frac{1}{\sqrt{2}}\frac{\lambda}{|\lambda|}$$
(29)

Further, from Eq. (24),

$$q = |\lambda|;$$
(30)

$$E_1 = E_{atom}[1 - |\lambda|];$$
(31)

$$E_3 = E_{atom}[1 + |\lambda|]$$
(32)

Then, Eq.(26) simplifies to the result

$$|\psi_\alpha(t)> = \frac{1}{2}\exp(-iE_{atom}t/\hbar)\left\{-\frac{\lambda}{|\lambda|}e^{iE_{atom}|\lambda|t/\hbar}[|\psi_1> + |\psi_2>] + e^{-iE_{atom}|\lambda|t/\hbar}[|\psi_3> + |\psi_4>]\right\}$$
(33)

To interpret the time changing nature of $|\psi_\alpha(t)>$, we expand the two square brackets in Equation (30) in the basis $|\Phi_k>$, k=1, 2, 3, 4. So, we can then write

$$|\psi_1> + |\psi_2> = \sum_{k=1}^{4} f_k |\Phi_k>$$

(34)

with

$$f_1 = f_2 = -\frac{1}{\sqrt{2}} \frac{\lambda}{|\lambda|} \ldots\ldots\ldots\ldots\ldots\ldots\ldots (a);$$

$$f_3 = f_4 = \frac{1}{\sqrt{2}} \ldots\ldots\ldots\ldots\ldots\ldots\ldots\ldots (b)$$

(35)

Also

$$|\psi_3> + |\psi_4> = \sum_{k=1}^{4} g_k |\Phi_k>$$

(36)

with

$$g_1 = g_2 = \frac{1}{\sqrt{2}}; \ldots\ldots\ldots\ldots\ldots\ldots\ldots (a)$$

$$g_3 = g_4 = \frac{1}{\sqrt{2}} \frac{\lambda}{|\lambda|} \ldots\ldots\ldots\ldots\ldots (b)$$

(37)

Under these conditions, Eq. (33) can be simplified to be

$$e^{iE_{atom}t/\hbar}|\psi_\alpha(t)> = \frac{(|\Phi_1> + |\Phi_2>)}{2\sqrt{2}} \left[ e^{iE_{atom}|\lambda|t/\hbar} + e^{-iE_{atom}|\lambda|t/\hbar} \right]$$

$$- \frac{\lambda}{|\lambda|} \frac{|\Phi_3> + |\Phi_4>}{2\sqrt{2}} \left[ e^{iE_{atom}|\lambda|t/\hbar} - e^{-iE_{atom}|\lambda|t/\hbar} \right]$$

(38)

which can be represented in angular form as

$$|\psi_\alpha(t)> = \frac{e^{-iE_{atom}t/\hbar}}{\sqrt{2}} \left\{ \cos\left(\frac{E_{atom}|\lambda|t}{\hbar}\right)(|\Phi_1> + |\Phi_2>) - i\sin\left(\frac{E_{atom}|\lambda|t}{\hbar}\right) \frac{\lambda}{|\lambda|}(|\Phi_3> + |\Phi_4>) \right\}$$

(39)

Eq.(39) shows entangled states similar to those in Eq.(7) in PC. However, our result is more general as we have included the non-interacting part of the energy in our model. Moreover, our results could further be generalized to n-particle states also. Hence the Eq.(33) can still be further generalized for the case of nonzero detuning.

**Non-Resonant Subcase ( $\varepsilon \neq 0$ )**

In general, the energies of the photon mode will not exactly match the energy difference between the atomic ground state and the atomic excited state. That is, there is some detuning and $\varepsilon \neq 0$. Then it is straightforward to extend the results of Part C to allow detuning. The expansions shown in Eqs.(34) and (36) remain valid, but the coefficients $f_k$ and $g_k$ can be easily generalized to the results

$$f_1 = f_2 = -\sin\theta; \qquad (a)$$
$$f_3 = f_4 = \cos\theta; \qquad (b) \qquad (40)$$

and

$$g_1 = g_2 = \cos\theta; \qquad (a)$$
$$g_3 = g_4 = \sin\theta. \qquad (b) \qquad (41)$$

Finally, Eq.(39) is replaced by the more general result

$$|\psi_\alpha(t)\rangle = e^{-iE'_{atom}t/h}\{F(q,\theta,t)|\psi_\alpha\rangle + G(q,\theta,t)|\psi_\beta\rangle\}. \qquad (42)$$

In Eq.(42), we have employed the definitions

$$E'_{atom} = E_{atom}(1 + \frac{3}{2}\varepsilon); \qquad (a)$$
$$F(q,\theta,t) = \cos(qE_{atom}t/h) - i(\cos 2\theta)\sin(qE_{atom}t/h); \qquad (b) \qquad (43)$$
$$G(q,\theta,t) = -i(\sin 2\theta)\sin(qE_{atom}t/h). \qquad (c)$$

Also note that

$$|\psi_\alpha\rangle = |\psi_\alpha(0)\rangle = \frac{1}{\sqrt{2}}\{|\Phi_1\rangle + |\Phi_2\rangle\}; \qquad (a)$$

$$|\psi_\beta\rangle = \frac{1}{\sqrt{2}}\{|\Phi_3\rangle + |\Phi_4\rangle\}. \qquad (b) \qquad (44)$$

In Section IV that follows, we will discuss the new effects present for entanglement exhibited by Eq.(42), as contrasted with the resonant case result of Eq.(39)

**Comparison of PC Results with JC Model Results**

To compare the result of Eq.(39) with the results of PC, we compare their notation with ours with the following correspondence

$$|0\rangle|-\rangle|-\rangle = |G\rangle ..........................(a)$$
$$a^+ = \frac{1}{\sqrt{2}}(a_A^+ + a_B^+)..............................(b)$$
$$(a^+|0\rangle)|-\rangle|-\rangle = \frac{1}{\sqrt{2}}(|\Phi_1\rangle + |\Phi_2\rangle)................(c) \qquad (45)$$
$$|0\rangle|+\rangle|-\rangle = |\Phi_3\rangle ........................(d)$$
$$|0\rangle|-\rangle|+\rangle = |\Phi_4\rangle ........................(e)$$

In the study of PC, the authors set

$$\lambda = 1 = \frac{E_{atom}}{h}$$

Comparison shows agreement of our results with those of PC (for positive $\lambda$) and with their equation (7). However, we can see that

(a) The oscillating factor $e^{-iE_{atom}t/\hbar}$ of Eq.(39) is missing in Eq.(7) of PC. This is because they are assuming that the complete Hamiltonian includes the interaction term which couples the atom to the photon modes.

(b) Eq.(7) of PC does not contain the factor -i in the second term of our Eq.(39). However our results include PC's results as a special case.

We can bring agreement between the second term of Eq.(39) and the second term of Eq.(7) of PC by the following: Replace the excited state of atomic wave functions for both atoms that appear in the work of PC by $-i\Phi_{j^+}$ (PC). Here, $\Phi_{j^+}$ (PC) denotes the excited atomic wave functions as used by PC. (This replacement is simply a multiplication by a phase factor and hence is an equally valid representation of the excited states). It then follows that the second terms of our Eq.(39) and Eq.(7) of PC are now identical.

## Discussions

We have studied entanglement in a two-atoms with two-photon modes system in the rotating wave approximation in the JC model [following HN], using the second quantization, as was used by PC. However, PC has studied only the resonant case ($\varepsilon = 0$), which is deployed by Eq.(39). We use the Hamiltonian proposed by HN for JC model for the two-atom and two-photon system and the second quantization. At time t=0, only the first term of Eq.(39) is non-zero. At this time, both atoms are in their ground state (see Eq. (18)). However, at the later time $t=t_0 = \frac{\pi}{2}(\frac{h}{|\lambda|E_{atom}})$, only the second term is present. Now, the following entanglement statement can be deduced from Eqs. (17) and (33): If atom A is in a ground state, then we know, with certainty, that atom B is excited, and vice versa. In other words, it is impossible for both atoms to be in their ground states, at time $t=t_0$.

There is a periodic increase and decrease of the strength of the entanglement, as expressed by Eq.(39). The period for a full cycle is ($T = \frac{2\pi h}{|\lambda|E_{atom}}$). Now turn to the more general non-resonant case, expressed by Eq. (42), ($\varepsilon \neq 0$). This highlights the interesting fact that the period of entanglement is a function of atomic energy. With the increase in atomic energy, the time period will decrease and vice a versa. The dimensionless parameter λ has a similar effect on the time period. However, since λ is a ratio between two types of energies, the main parameter can be considered as $|\lambda|E_{atom}$, that is inversely proportional to T and can control the time period of entanglement.

The statements in the preceding paragraphs are modified by noting that maximum entanglement occurs at time $t = t_0' = (\frac{\pi h}{2qE_{atom}})$. From Eq.(24), $|\lambda| \leq q$, for $\varepsilon \neq 0$. Hence, $t \leq t_0$, when $\varepsilon \neq 0$, maximum entanglement occurs more quickly than in the resonant case. The period of oscillation is now $T = (\frac{2\pi h}{qE_{atom}})$.

To summarize, we have studied the entanglement of two atoms and two photon states and its time evolution. However, we have entered into a model (Jaynes-Cummings) that can be extended to a larger collection of atoms in the presence of a larger number of photon modes.

Moreover, using the modified form of our master equation, we can calculate the dissipation of any energy mode from a cavity. The master equation gives the major source for the dissipation of photon energy, but, it does not account for the contribution to loss due to the interaction of the atoms with the cavity. This dissipative dynamics of cavity can be derived from the leakage of cavity photons due to the imperfect reflectivity of the cavity mirrors. It is usually considered in the JC model that the presence of atoms in a cavity may not significantly affect [7] the cavity losses. Due to the possibility of entanglement, it may no longer be true. We still have to find out that how the entanglement can be maintained and now the entangled states could still be handled individually.

It is also worth-mentioning that we are not the only one using the JC model. Some of the other papers [10-12] have also studied entanglement in JC model. Though, our model of two atoms and two photons is not used previously.